%% file: project.tex
\documentclass[10pt,conference]{IEEEtran}

\usepackage{times,amsmath,epsfig}
\usepackage{type1cm}     
\usepackage{algorithm}     
\usepackage{algorithmic}     
\usepackage{graphicx}     
\usepackage{epstopdf}
\usepackage{xspace}     
\usepackage{balance}     
\usepackage{booktabs}     
\usepackage[bf,tableposition=top]{caption}     
\usepackage[hyphens]{url}     
\usepackage[bookmarks, pdftex, colorlinks=false]{hyperref}     
\usepackage[square,numbers]{natbib}     
\usepackage[font=small,labelfont=scriptsize]{caption}
\usepackage[section]{placeins}

\usepackage{amsmath,amssymb}

\title{Socially-Aware Distributed Hash Tables for Decentralized Online Social Networks}

\author{
{
Muhammad Anis Uddin Nasir{\small $^{\#1}$},
Sarunas Girdzijauskas{\small $^{\#2}$}
Nicolas Kourtellis{\small $^{*3}$}
}
\vspace{1.6mm}\\
\fontsize{10}{10}\selectfont\itshape
$^{\#}$KTH Royal Institute of Technology, Stockholm, Sweden\\
$^{*}$Telefonica Research, Barcelona, Spain\\
\fontsize{9}{9}\selectfont\ttfamily\upshape

$^{1}$anisu@kth.se,
$^{2}$sarunasg@kth.se,
$^{3}$nicolas.kourtellis@telefonica.com,
}

\begin{document}

\maketitle

\begin{abstract}
Many decentralized online social networks (DOSNs) have been proposed due to an increase in awareness related to privacy and scalability issues in centralized social networks. 
Such decentralized networks transfer processing and storage functionalities from the service providers towards the end users.
DOSNs require individualistic implementation for services, (i.e., search, information dissemination, storage, and publish/subscribe). 
However, many of these services mostly perform social queries, where OSN users are interested in accessing information of their friends.
In our work, we design a socially-aware distributed hash table (DHTs) for efficient implementation of DOSNs.
In particular, we propose a gossip-based algorithm to place users in a DHT, while maximizing the social awareness among them. 
Through a set of experiments, we show that our approach reduces the lookup latency by almost 30\% and improves the reliability of the communication by nearly 10\% via trusted contacts.

  
\end{abstract}

\input{intro}
\input{overview}
\input{rel-work}

\input{pitch}

\input{problem-def}
\input{solution}
\input{evaluation}
\input{discussion}

\bibliographystyle{IEEEtran}
\bibliography{IEEEabrv,biblio}

\end{document}

%% file: intro.tex
\section{Introduction}


Online social networks are ubiquitous, from friendship networks like Facebook, to professional networks like LinkedIn.
A variety of different services are supported in these platforms, such as search, information dissemination, storage, profile management, and application integration.
Such networks are well known to have small world properties, i.e., high clustering and small diameter~\cite{benevenuto2009characterizing}.

Currently, most of the social networks operate in a centralized fashion with a central service responsible for providing the social network services.
The incentive for a provider is the access to large amounts of data, which can be used for business-related purposes~\cite{debatin2009facebook, dwyer2011privacy}.
However, these incentives have raised privacy concerns among users. 
Therefore, in the last decade, researchers and the open source community have proposed various decentralized solutions (e.g., ~\cite{koll2014soup,nilizadeh2012cachet, kapanipathi2011privacy,buchegger2009peerson,cutillo2009safebook,kourtellis2010prometheus}) that remove dependency on a centralized provider.

Indeed, a decentralized environment, such as the one in the aforementioned solutions, requires independent implementation for each of these components for good quality of service.
For example, for information dissemination, social users are interested in propagating the information to their direct friends, whereas for search service, they are interested in accessing the complete knowledge of the network.
However, individualistic employment for different services imposes an expensive and challenging burden on a DOSN community. 

DHTs~\cite{lua2005survey} are a very promising solution for DOSNs since they provide all required functionalities with a limited peer degree in the resulting overlays~\cite{kourtellis2010prometheus,jahid2012decent, shakimov2011vis}.
Such overlays are small-world in nature and have efficient routing properties. 
However, current DHTs create such overlays solely based on the peer IDs which are assigned uniformly at random and do not reflect the social graph structure of DOSNs. 
This significantly downgrades the performance of DHT-based DOSNs since most of the workloads directly reflect the topology of the social graph mapped on the overlay~\cite{6494568}. 
On a DHT-based DOSN, such requests would correspond to generating expensive relay traffic for performing simple actions on each social link.


\begin{figure}[t]
\begin{center}
\includegraphics[width=\columnwidth]{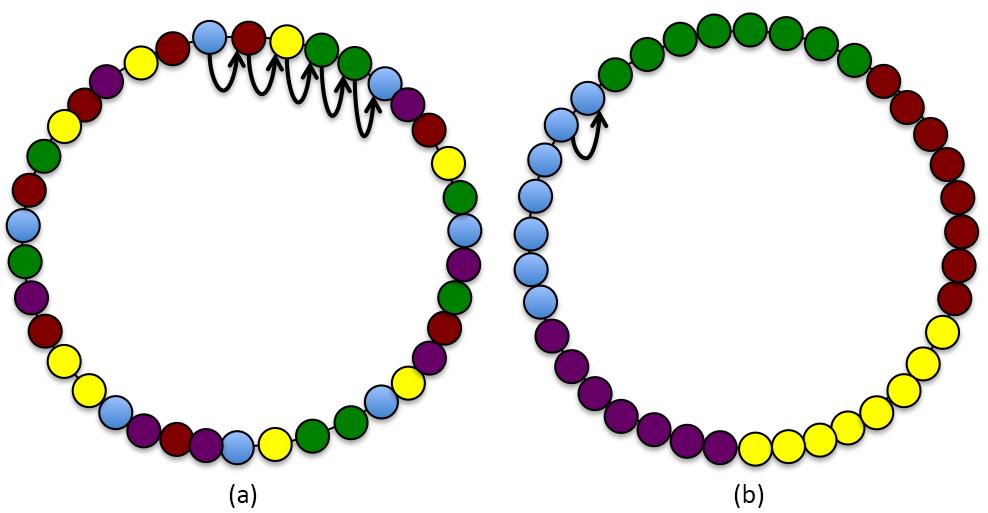}
\caption{Example of creating an overlay. The edges (in black color) represent a search request from one blue node to another blue node in the overlay. (a) search route in a random DHT-based overlay, and (b) search route in a socially-aware DHT overlay.}
\label{fig:toy-example}
\end{center}
\end{figure}

DOSN users can be arranged in a friend-to-friend network with each user maintaining the information related to their direct connections.
Each user leverages a push mechanism to send an update to all of their neighbors. 
DOSNs like PeerSON \cite{buchegger2009peerson} propose a similar solution with an additional lookup service. 
Such solutions are easy to implement as they involve only interested nodes for information distribution, which leads to a low data propagation time.
However, they create various challenges, like work imbalance and scalability, due to the presence of skew in small-world social networks.
For instance, a scheme may impose heavy workload on high degree users, which makes the service unusable for such users operating via a hand-held device.

Figure \ref{fig:toy-example}  shows a toy example of a search request in two different overlays: a) a random DHT-based overlay and b) a socially-aware DHT-based overlay. 
The overlay Figure \ref{fig:toy-example} (a) is a simple arrangement of users in the form of a ring, where users are assigned identifiers uniformly at random.
This type of assignment will lead to high communication cost, as many uninterested peers (relay nodes) are involved in the search process.
However, search process improves by placing close socially-connected users (same color) in the ring, as shown in Figure \ref{fig:toy-example}(b).



In our work, we aim to improve the performance of DHT-based DOSN services by designing a socially-aware DHT.
Specifically, we leverage a DHT as an underlying overlay and achieve social awareness by migrating nodes across the DHT, without modifying the actual overlay.
Since it is known that social graphs are small-world in nature, our task would be to identify proper subsets of our social network graph which most closely represent the graphs observed in DHT overlays. 
In particular, we need to assign IDs for each participating node in such a way, that a decentralized greedy routing algorithm would perform most efficiently while traversing  mostly existing social-links.
Also, this social awareness will result in improving the reliability among nodes in the overlay~\cite{marti2005dht} as messages between them will traverse friendly nodes.

We propose to smartly embed social networks on top of DHTs, like symphony \cite{manku2003symphony}. 
Our work is inspired by \cite{sandberg2006distributed}, where an approach was proposed for embedding small world graph on top of a Kleinberg's two dimensional grid using a statistical estimation. 
Following the same path, we propose to embed social network on top of small world DHTs.
We believe that the presence of small-world properties, i.e., high clustering and small diameter, in both social networks and DHTs, will enable direct embedding of social networks on a DHT.
Small world properties allow us to create a notion of ties among nodes in both networks.
This information can further be used to create mapping of users on both networks.  

We propose a gossip-based algorithm to perform this embedding of the social network in the DHT overlay.
We show that our approach reduces the lookup latency by almost 30\% in the network and improves the reliability of the communication by nearly 10\% via trusted contacts.


%% file: overview.tex
\section{Overview and Related Work}\label{sec:overview}

In this section we outline the steps of our method and provide a comparison with related work on DOSNs built on top of DHTs.

\subsection{Overview}

The problem we are tackling is similar to clustering, where we need to arrange nodes in an overlay while taking into account their social proximity.
One solution to the problem is taking the snapshot of the social network, running a clustering algorithm, and assigning the identifiers to nodes based on their clusters.
However, this can be a costly operation, possibly depended on acquiring a global view of the social graph.
On the contrary, we propose the use of an incremental and decentralized gossip-based algorithm for embedding the social network on top of a DHT.
The gossip-based approach towards grouping similar nodes in our work is inspired by \cite{ jelasity2009t, voulgaris2005epidemic, gupta2003kelips}.
In our work, we use the gossip-based approach to optimize for social proximity of nodes in the graph.
The algorithm works in two phases.

In the first phase, or \emph{initialization}, the algorithm randomly initializes the DHT overlay, without taking into account the social structure.
In the second phase, or \emph{refinement}, the algorithm aims at migrating the nodes in the overlay closer to their social friends.
There are various definitions of closeness in an overlay, such as the euclidean distance in the id space, and the number of hops in the overlay.
In our work, we define a utility function that can use either of the two definitions, i.e., euclidean distance or hop count.
However, our algorithm is capable of adapting to any other definition of closeness, like round trip time or geographical proximity~\cite{jelasity2009t}.
Similarly, there are different ways to rank the social ties of users, e.g., mutual friends (triangle count)~\cite{prat2012shaping}, user interactions~\cite{benevenuto2009characterizing}, and others.
In our work, we select the triangle count as the measure for estimating the strength of ties.
We select this metric due to its simplicity, adaptability and decentralized nature.


The second phase of our algorithm runs in multiple iterations. 
In each iteration, every node tries to maximize its social awareness by performing identifier swapping with a selected node via gossiping~\cite{nasir2014gossip}.
For peer selection, we compare various decentralized schemes, i.e., random, direct, greedy and smart (see section \ref{sec:solution}).
The peer selection schemes (inspired from~\cite{jelasity2009t, voulgaris2005epidemic}) are configured to take into account the social ties.
Based on an empirical study, we find the direct peer selection scheme as a good option for selecting nodes to swap ids.
In this scheme, a node asks a direct social friend to return his overlay finger (link/connection) for swapping.
Due to its decentralized nature, identifier swapping process can be performed in a massively parallel way and requires only local knowledge of graph topology (i.e., every node knows its direct neighbors in the overlay). 

After peer selection, the node evaluates the cost of swapping its identifier with the identifier of the selected node, by looking at its social proximity.
A node adapts to a new identifier, if the candidate's identifier brings it closer to its social friends.
The algorithm groups similar (connected) users together in the overlay. 
Therefore, nodes in a social network require fewer steps to communicate between each other. 
Moreover, once connected nodes are grouped together, they tend to stay in the same neighborhood in the overlay. 
This behavior accelerates the convergence process and reduces the number of identifier swaps between nodes.
However, nodes in the overlay require to periodically execute the algorithm, to keep the overlay consistent and navigable, due to nodes joining and leaving the social network. 

%% file: rel-work.tex
\subsection{Related Work}\label{sec:rel-work}
The idea of socially-aware overlays has been previously proposed in designing efficient systems while using an underlying social graph structure~\cite{liu2009locality,ciullo2010network,kourtellis2015special}.
Similarly, there are past works proposing to improve locality-awareness in a system in order to achieve better network latency \cite{qiu2007towards,resmi2015fluidify}.
However, OSNs are typically used for social queries, where performance is highly improved by increasing social awareness \cite{kourtellis2015special, kourtellis2010prometheus}. 

Further, for a DOSN, a n{\"a}ive way to create a social overlay is to use a structured overlay like DHT~\cite{jahid2012decent, shakimov2011vis}. 
However, these solutions lack social awareness, which leads to higher communication overhead and less reliability~\cite{marti2005dht}. 
Optimization like SPROUT~\cite{marti2005dht} improves the efficiency and reliability of structured overlays for DOSNs.

Arranging social users in a friend-to-friend network is also an attractive choice for implementing a DOSN~\cite{koll2014soup,nilizadeh2012cachet,buchegger2009peerson,shakimov2009privacy, seong2010prpl}, and also in \cite{kourtellis2015special} using the primary communication layer for social inferences.
However, such services require each user to maintain connections with all their friends, which makes this approach unscalable and impractical for handheld devices.
Moreover, the overlay maintenance overhead can be reduced by using an external lookup service, similar to~\cite{buchegger2009peerson}.
Nevertheless, such an external service is a single point of failure and can lead to privacy and security issues.

Safebook~\cite{cutillo2009safebook} leverages the social trust between users by building a network of trusted peers that store OSN data.
To increase system reliability and availability, each user's data are replicated on trusted friends of this user.
This scheme is not scalable due to the overlay maintenance overhead.

Diaspora~\cite{bielenberg2012growth} is a super-peer based architecture, with network of independent, federated servers that are administrated by individual users. 
Supernova~\cite{sharma2012supernova}, proposes a similar service using a super-peer based approach for DOSNs, where super peers arrange themselves to provide lookup services for other users.
Such networks transfer the problem of privacy from a central provider to anonymous servers, which increases the risk of security breaches. 



%% file: pitch.tex
\section{Preliminaries}
\label{sec:motivation}

In order to design an overlay for DOSNs, one of the major requirement is having a limited or bounded number of connections per user (especially for hand-held devices), while providing guarantees of deterministic and bounded reach.
For this particular reason, a DHT is one of the most optimal choices among all the other possible solutions~\cite{buchegger2009peerson, kourtellis2010prometheus, shakimov2011vis, cutillo2009safebook, bielenberg2012growth}.

\subsection{Distributed Hash Tables and Symphony}

Distributed hash tables is a class of distributed systems that can provide autonomous, efficient, fault-tolerant and scalable lookup service. 
DHTs have been used in the past to provide various services, like distributed file systems, distributed storage, domain name systems, web caching, file sharing and content distribution systems~\cite{lakshman2010cassandra,cohen2008bittorrent,oram2001peer}.
We use a DHT to provide an overlay for decentralized online social networks.
Information in a DHT is stored in the form of a key-value pair.
Any participating node in a DHT efficiently retrieves a value associated with a given key.
We select symphony~\cite{manku2003symphony}, due to the presence of small world characteristics \cite{kleinberg2000small}.


In Symphony, nodes arrange themselves in a ring structure and are assigned an identifier from a uniform identifier space between (0,1].
Each node in Symphony maintains two links to their immediate neighbors (predecessor and successor).
Moreover, each node maintains $k$ additional long range links to improve the lookup process.
Long range links are created using a probability distribution function, which guarantees the presence of small world properties, i.e., large number of short links and small number of long link.
Symphony is a flexible, fault tolerant, stable lookup service that provides lookup with an average latency of O($\frac{1}{k}\log N)$ hops.

Chord~\cite{stoica2001chord} is a special case of Symphony.
In Chord, nodes are arranged in the form of a ring, where each node is assigned an identifier from an identifier space between (0,1].
Each node maintains fingers (pointers) only for \begin{math}O(logN)\end{math} other nodes in the network. 
In Chord, long range links are created using a function $f(i)=2^i$, where $i \in 1, \ldots, \log N$.
Chord guarantees lookup for each node in the network in $O(\log N)$ steps.

\subsection{Embedding Social Networks via DHT Finger Rewiring}

The motivation of our work relies on the fact that random assignment of social users on a DHT does not guarantee any social awareness~\cite{6494568}.
The lack of social awareness results in the participation of many potentially uninterested users (relay nodes) in the lookup process, which leads to higher network cost and increased security risks.

Sandberg~\cite{sandberg2006distributed} proposed an approach for embedding small world graphs on top of a Kleinberg's two dimensional grid using a statistical estimation. 
It is a well known fact that most of the online social networks follow the properties of small world networks~\cite{benevenuto2009characterizing}.
Therefore, we expect that there can be an embedding of a social graph on an overlay like Symphony that can improve lookup performance and reliability, without jeopardizing the guarantees offered by the DHT.


One of the simple solutions that improves the social awareness of a DHT is updating the finger table of each node in the DHT to point to social connections.
In Symphony, this can be achieved by simple modification of the sampling function for long distance links.
Concretely, rather than sampling and selecting a random peer, we need to sample and select a friend (social connection) closest to that peer. 
This simple modification drastically improves the social awareness compared to the random symphony overlay, as nodes directly point to their friends. 
Moreover, this approach maintains the small world nature of the Symphony overlay, thus offering similar bounds on the lookup calls. 
However, this greedy heuristic does not provide an optimal solution, as each user in the overlay takes a local decision of keeping their immediate friends in their routing table.


%% file: problem-def.tex

\section{Problem Definition}
\label{prob-def}
We formulate our problem by considering an undirected social graph \begin{math}G=(V,E)\end{math}, where \begin{math}V\end{math} is the set of vertices and \begin{math}E\end{math} is the set of edges, connecting the vertices. 
The graph contains $|V|$ nodes and $|E|$ edges, where:
\begin{center}
$E = \{e_{ij} = (i,j): i \rightarrow j \text{ and } j\rightarrow i\}$
\end{center}

Each node \textit{i} in a graph has a neighborhood set of neighbors defined as:
\begin{center}
$N_i = \{j\in V: e_{ij}\in E\}$
\end{center}
The graph, G, has a static, N $\times$ N symmetric adjacency matrix A, where:
\[ A_{ij} = \left\{ 
  \begin{array}{l l}
    1 & \quad \text{if $e_{ij} \in E$}\\
    0 & \quad \text{if $e_{ij} \notin E$}
  \end{array} \right.\]

Further, we define the strength of ties among users in a social graph.
We leverage the notion of number of triangles that two nodes share in the graph to find the strength among the two nodes.
A triangle in a graph is similar to a mutual or common friend in a social network.
Therefore, we define the strength between two nodes $i$ and $j$ as follows:
\begin{equation}
s_{ij} = \frac{\{|N_i \cap N_j|: i,j \in V\}}{|N_i|}
\end{equation}

Along with the social graph, each vertex $v\in V$ from the graph \begin{math}G\end{math} also participates in an overlay (symphony), where each node is assigned an identifier from an identifier space (0,1]. 
This means that each node in the social graph has one-to-one mapping to an identifier in the symphony overlay.
Further, nodes create fingers depending on the algorithm of the overlay.
For example, in symphony~\cite{manku2003symphony}, each node creates a set of fingers to its immediate neighbors and creates long range links using the probability distribution function. 

For a DHT, we define the distance between two nodes $i$ and $j$ as $d_{ij}$.
We define two different distance metrics: 1) euclidean distance between the two node ids, and 2) lookup latency (number of hops in the overlay) between the two nodes.

As each node gets an identifier from a 1-dimensional space, the euclidean distance between node $i$ and node $j$ with identifiers $x_i$ and $x_j$, respectively, is given by the absolute distance:
\begin{center}
$d_{ij}=|x_i-x_j|$
\end{center}

The lookup latency between the two nodes in the DHT overlay is calculated using the symphony routing algorithm.
The distance $d_{ij}$ in this case is equal to the number of intermediary nodes a node $i$ needs to traverse to reach the destination node $j$.
Observe that the social strength between users is calculated using the social graph, whereas the distance between two nodes is extracted from the information about their identifiers in the overlay.

\subsection{Utility Function}

We represent the problem as an embedding problem, where we map the vertices in the social graph to DHT nodes.
To tackle the problem, each node ranks their neighbors based on the strength of social ties.
Also, each node needs to know the information about short range and long range links.

We define two cost functions for our problem.
Both cost functions use the same minimization objective. 
However, they differ in the definition of the distance between nodes, which is parameter $d_{ij}$:
\begin{equation}
\min \sum\limits_{i}\sum\limits_{j} s_{ij} \times d_{ij}
\end{equation}

The first parameter of the cost function $s_{ij}$ is the strength of ties between two nodes in a social graph.
The second parameter of the cost function  $d_{ij}$ is the distance between two nodes in the DHT overlay.
The cost function aims at minimizing the distance between strongly connected social users.

At any time $t$, a node $i$ calculates the cost $C_i(t)$ to swap node ids with another node, given its neighborhood $N_i$:
\begin{equation}\label{eq:cost-function}
C_i(t) = \sum_{k \in N_i} s_{ik} \times d_{ik}(t)
\end{equation}

This cost leverages the parameter $s_{ik}$, which is a fixed parameter that does not change overtime t, and the parameter $d_{ik}(t)$ which changes overtime, depending on the current distance of nodes with their friends on the identifier space at time $t$.

%% file: solution.tex
\section{Algorithm}
\label{sec:solution}

To solve the aforementioned problem, we utilize a gossip-based algorithm proposed by~\cite{nasir2014gossip}.
The algorithm we propose works in two phases: a) Initialization and b) Refinement.

\subsection{Initialization}

To initialize the algorithm, we begin by taking the input in the form of a social graph.
Each node in the graph is randomly assigned an identifier from an identifier space (0,1].
Further, these nodes create a DHT overlay by constructing pointers to their immediate neighbors and by creating long range links to other nodes in the overlay, following the symphony protocol~\cite{manku2003symphony}.

This phase distributes the nodes randomly in the overlay.
The random distribution of connected nodes place them uniformly in the identifier space, resulting in lack of social awareness.
Hence, each node in the symphony overlay traverses random nodes (relay nodes) in order to reach another node, thus allowing two social friends to communicate (in the application layer).

\subsection{Refinement}

Refinement aims to improve social awareness in the overlay by continuously moving the nodes in the overlay closer to their friends.
At each time instance $t$, each node locally computes the strength and distance with other nodes and makes local decisions if to swap his id with any of them or not.
The benefit of the local computation is two-fold. 
First, it enables our algorithm to work in a distributed manner, as it does not require a global view of the graph.
Second, each node greedily improves the social awareness by performing lightweight operations.

Refinement works in three steps: a) node selection, b) cost evaluation, and c) identifier exchange.
First, each node selects another node from the network under a particular selection scheme.
Observe that each node has access to two types of nodes: social links and overlay links.
Social links are his friends in the social graph, whereas overlay links are the immediate neighbors in the overlay.
We investigate four different approaches for node selection, where a node $i$ first selects another node $m$ who provides a finger $j$ from his table, as a candidate to swap ids with $i$:

\textbf{Direct:} A node $i$ selects one of his friends in the social graph uniformly at random to be the node $m$.

\textbf{Greedy:} A node $i$ selects its friend with strongest tie as node $m$.

\textbf{Smart:} A node $i$ selects a node $m$ uniformly at random from its top k strongest friends.

After node $m$ is selected, $m$ selects a node $j$ that resides in its finger table.
This node $j$ will be the one to be considered for swapping with $i$ if the cost function allows it, as explained next.
The intuition behind this approach is that by selecting a finger of direct neighbor $m$, at each time $t$, node $i$ moves closer to its social friends.
We also compare with a random selection scheme as a baseline.

\textbf{Random:} A node $i$ selects a random other node $j$ from the graph to swap ids.
This scheme can be implemented using a peer sampling service, similar to \cite{jelasity2007gossip}.

Second, the algorithm evaluates the cost of adapting the identifier of node $i$ to node $j$.
Each node uses the cost function defined in Eq.~\ref{eq:cost-function} to calculate the cost of the new identifier at time interval $t$.
In particular, the algorithm calculates the cost for both the nodes (local node and selected node) using their old and the candidate identifiers.

Suppose, we have two nodes $i$ and $j$ with their current cost at time t equal to $C_i(t)$ and $C_j(t)$, respectively.
We define the cost of node $i$ adapting to the identifier of node $j$ as $C_{ij}(t)$.
A node $i$, at any time instance $t$, evaluates the identifier adaption decision of changing its identifier to an identifier of node $j$ at the subsequent time instance $t+1$, using the following decision strategy:

\[
Cost(t) = C_{i}(t)+C_{j}(t)
\]
\[
Cost(t+1) = C_{ij}(t+1)+C_{ji}(t+1)
\]
\begin{equation}
\begin{cases}
i \text{ swaps ids with } j	& \quad \text{if } Cost(t) > Cost(t+1)	\\
\text{no swap}			& \quad \text{otherwise}			\\
\end{cases}
\end{equation}

Finally, the nodes exchange their identifiers if the exchange is resulting in minimizing the overall cost, and thus maximizing the social awareness in the system.


\subsection{Analysis}

We note that our algorithm executes in multiple rounds, where in each round, nodes try to perform identifier exchange to improve the social awareness in the system.
The algorithm operates greedily to optimize the cost.

The identifier exchange is a decentralized process, as each node is capable of performing it based on local information and by communicating with only one other node.
This enables parallel execution of the algorithm.
However, the algorithm needs to take care of conflicts: a node should only participate at most in one gossip phase (refinement) at a time.
In a completely decentralized environment, this can be achieved using a promise mechanism such as in the paxos protocol~\cite{lamport2001paxos}, where each node in a gossip phase, refuses any other incoming requests.

Due to the distributed nature of the protocol where nodes perform local optimizations of the cost function, there are no guarantees that the algorithm will lead to a steady state, and how close this steady state is to the state with a global minimum on the cost function.
However, as we demonstrate in our experimental evaluation, given real small-world graphs, the algorithm does converge to a steady state and this state has improved properties with respect to communication latency and reliability.

\subsection{Dealing with Failures}

Social overlays are sensitive to churn and might encounter non-negligible delays for various OSN services \cite{mega2012churn}.  
Therefore, a DOSN solution should successfully cope with different types of failures and churn.

Social graphs have been found to evolve in a linear fashion~\cite{kumar2010structure}.
This allows a DOSN overlay to adapt to evolving trends in the graph.
A hybrid overlay on top of a social graph requires any changes in the social graph to be reflected in the overlay.
Therefore, in order to make the overlay consistent and navigable, nodes in the overlay need to periodically update their pointers.

Similarly, instability in the overlay can be caused due to mobility or inactivity of peers, and peer or network failure.
Such temporary node failures can be handled by: a) using data replication on external data sources or other nodes in the network \cite{shakimov2009privacy}, b) providing guarantees for eventual consistency, and c) assuming socio-incentivized networks, where peers keep their computers online as much as possible to help their friends \cite{kourtellis2015special}.

Our approach naturally captures and adapts to the changes in the social graph through the gossiping process and appropriate cost functions associated with it. 
The temporary node failures are expected to be handled by the inherited techniques from the structured overlay maintenance schemes which our proposed technique has to be built on (e.g., ring stabilization of DHTs, etc.).

%% file: evaluation.tex
\section{Evaluation}\label{sec:evaluation}

In this section we describe the experimental evaluation performed to assess the gains of the proposed algorithm when compared to a symphony overlay.

\subsection{Experimental Setup}

We evaluate our proposed solution for socially-aware DHT through simulations with real datasets.

\subsubsection{Experimental Questions}

In our experimental evaluation we investigate different questions regarding the proposed algorithm:\\
\textbf{Q1:} What is the tuning cost of the algorithm?\\
\textbf{Q2:} How does node ordering impact algorithm convergence?\\
\textbf{Q3:} What are the performance gains with respect to lookup latency and reliability?\\
\textbf{Q4:} How robust is the algorithm to graph clustering?\\


\subsubsection{Algorithm Initialization}

We use Symphony as an underlying overlay and optimize the social awareness of the overlay using the gossip-based algorithm.
For symphony, we fix the $k=log(N)$, where $k$ is the number of long range links that each node maintains within the overlay~\cite{manku2003symphony}. 
This means that each node in the overlay maintains $2(1+k)$ connections (direct neighbors, $k$ outgoing links, and $k$ incoming links).

\subsubsection{Performance Metrics}

In our experiments we assess the performance gains of the algorithm by measuring three metrics:
1) latency in the overlay to reach a direct friend in the graph, expressed in average number of overlay hops, computed across all pairs of friends,
2) migration cost (the ratio of number of identifier exchanges over number of gossip attempts), and
3) reliability of the system with respect to how many friends a node can access directly in one hop~\cite{marti2005dht}.
We define the reliability of the algorithm in two ways: 1) how many direct friends occupy the finger table, and 2) the total number of n-hop friends in the finger table.

\subsubsection{Datasets}

\begin{figure}[t]
\begin{center}
\includegraphics[width=\columnwidth]{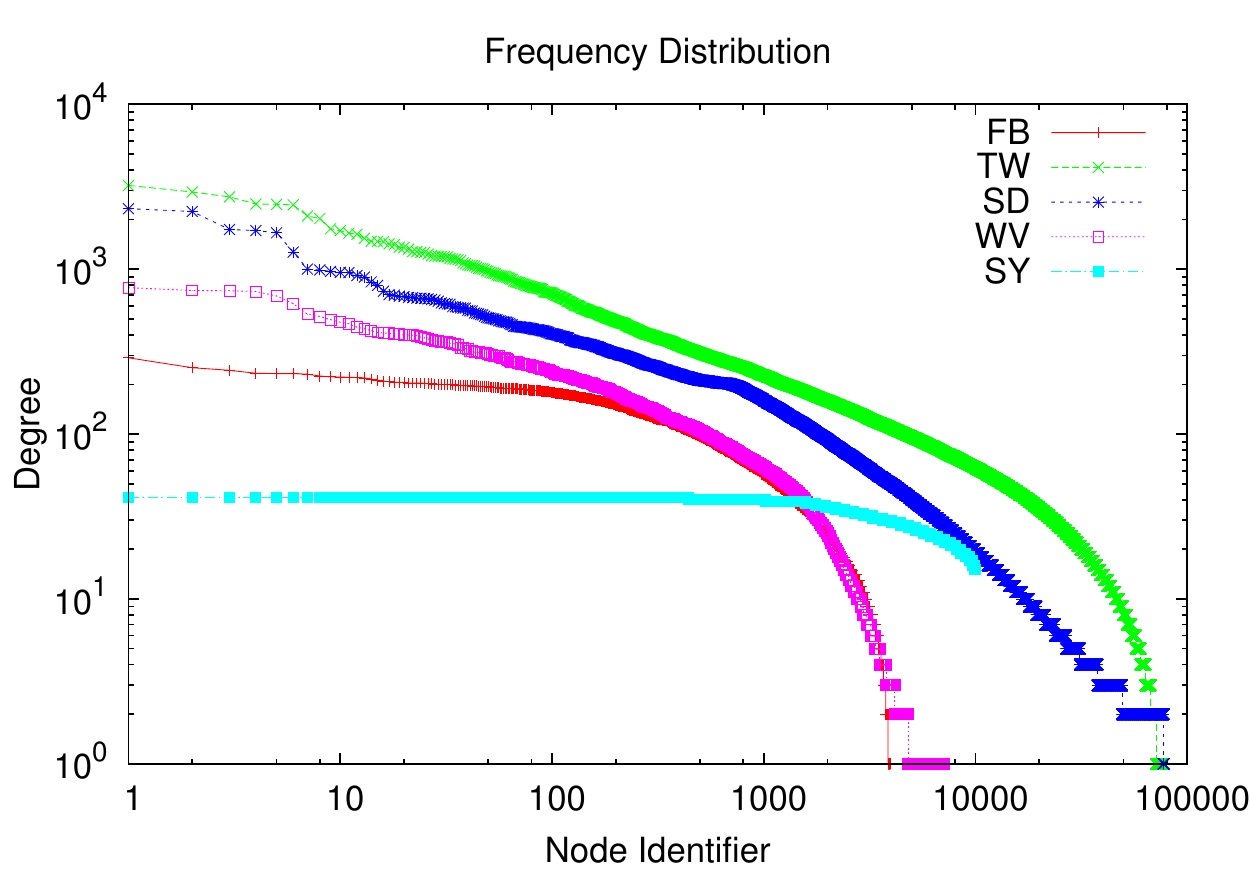}
\caption{Degree distribution of datasets used in the experiments.}
\label{fig:key-frequency-distribution}
\end{center}
\end{figure}

Table~\ref{tab:summary-datasets} shows the datasets used in the experiments.
These datasets\footnote{\url{http://snap.stanford.edu/data/index.html}} are chosen to impose real-world social networks link distributions on top of the overlay.
Figure~\ref{fig:key-frequency-distribution} shows the degree distribution of these social graphs.

\begin{table}[htbp]
\centering
\caption{Summary of the datasets used in the experiments. AD stands for average degree and ACC stands for average clustering coefficient.}
\small
\begin{tabular}{l c c c c c}
\toprule
Dataset		&	Symbol	&	Nodes	&	Edges	& 	AD	&	ACC \\ 
\midrule
Facebook		&	FB		&	4039		&	88234	&	42	&	0.6055 \\
Wiki-Vote		&	WV		&	7115		&	201524	&	28	&	0.1409 \\
Slashdot		&	SD		&	77360	&	1015667	&	13	&	0.0555 \\
Twitter		&	TW		&	81306	&	2484794	&	32	&	0.5653 \\
\midrule
Symphony	&	SY		&	10000	&	279959	&	28	&	0.043 \\
\bottomrule
\end{tabular}
\label{tab:summary-datasets}
\end{table}


\subsection{Experimental Results}
\label{experimental-results}

\textbf{Q1:} We perform experiments to evaluate the cost of tuning the algorithm.
In particular, we evaluate the cost of a) node selection, b) migration cost, and c) convergence time.

In the first experiment, we compare different ways to select a node.
We perform experiments comparing four node selection approaches that were discussed in section \ref{sec:solution}.
In this experiment, we use the FB dataset.
Initially, nodes are placed in the overlay, by randomly mapping social users to Symphony nodes.
Next, we execute the gossip algorithm, for 1000 iterations with the goal of minimizing the number of hops.
We observe the improvement in terms of lookup latency.

\begin{figure}[ht]
\begin{center}
\includegraphics[width=\columnwidth]{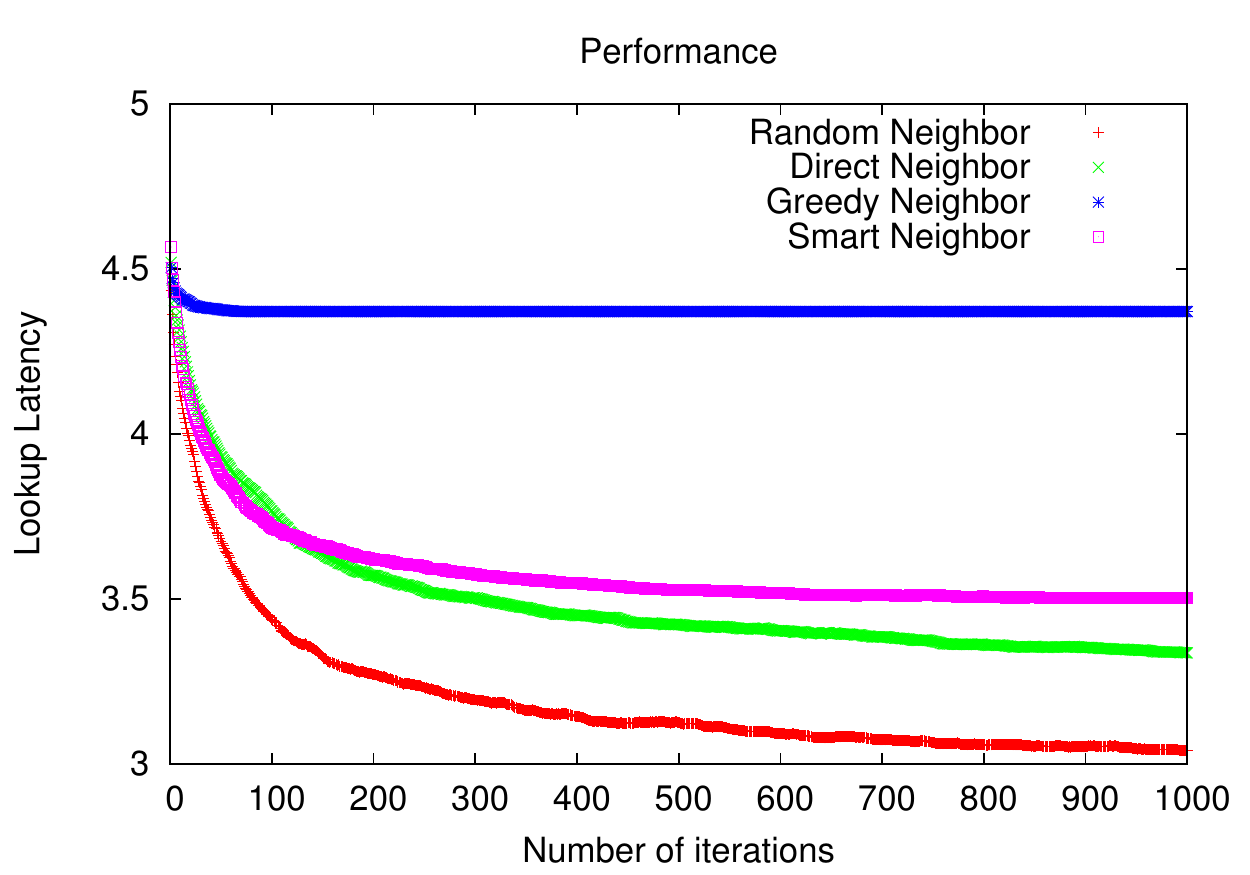}
\caption{Comparison of different node selection schemes (random, direct, greedy, smart) with the original symphony based overlay.
Lookup latencies are measured in terms of number of hops along the path to the destination node.
The cost function used in this experiment attempts to minimize the number of hops.}
\label{fig:peer-sampling-euc}
\end{center}
\end{figure}

Figure \ref{fig:peer-sampling-euc} shows the performance in terms of lookup latency for different node selection schemes.
The plot contains a label for symphony that represents the starting point of the algorithm.

Results show that the greedy node selection has the fastest convergence time.
However, it provides minimal improvements with respect to lookup latency.
This is due to the fact that in greedy selection, each node selects only one neighbor to gossip with and this reduced freedom has an impact in the overall converged state.

The other three approaches, i.e., random, direct and smart selection, achieve better improvements in the performance of the overlay.
However, random peer selection scheme is expected to lead to higher communication cost to find a node to start the gossiping phase, than the direct neighbor approach.
Therefore, we use the direct selection scheme for further experiments, as it provides nearly similar results as the random scheme but it affects only direct friends' node ids instead of random ids.

In the second experiment, we compare the migration cost of the keys.
Fraction of migration cost of the algorithm represents the number of identifiers exchanged between nodes over the attempted gossip actions. 
In this experiment, we only report the migration cost, when two nodes decide to swap their identifiers. 
Therefore, we ignore the cost of communication in case nodes decide not to swap their identifiers.

\begin{figure}[ht]
\begin{center}
\includegraphics[width=\columnwidth]{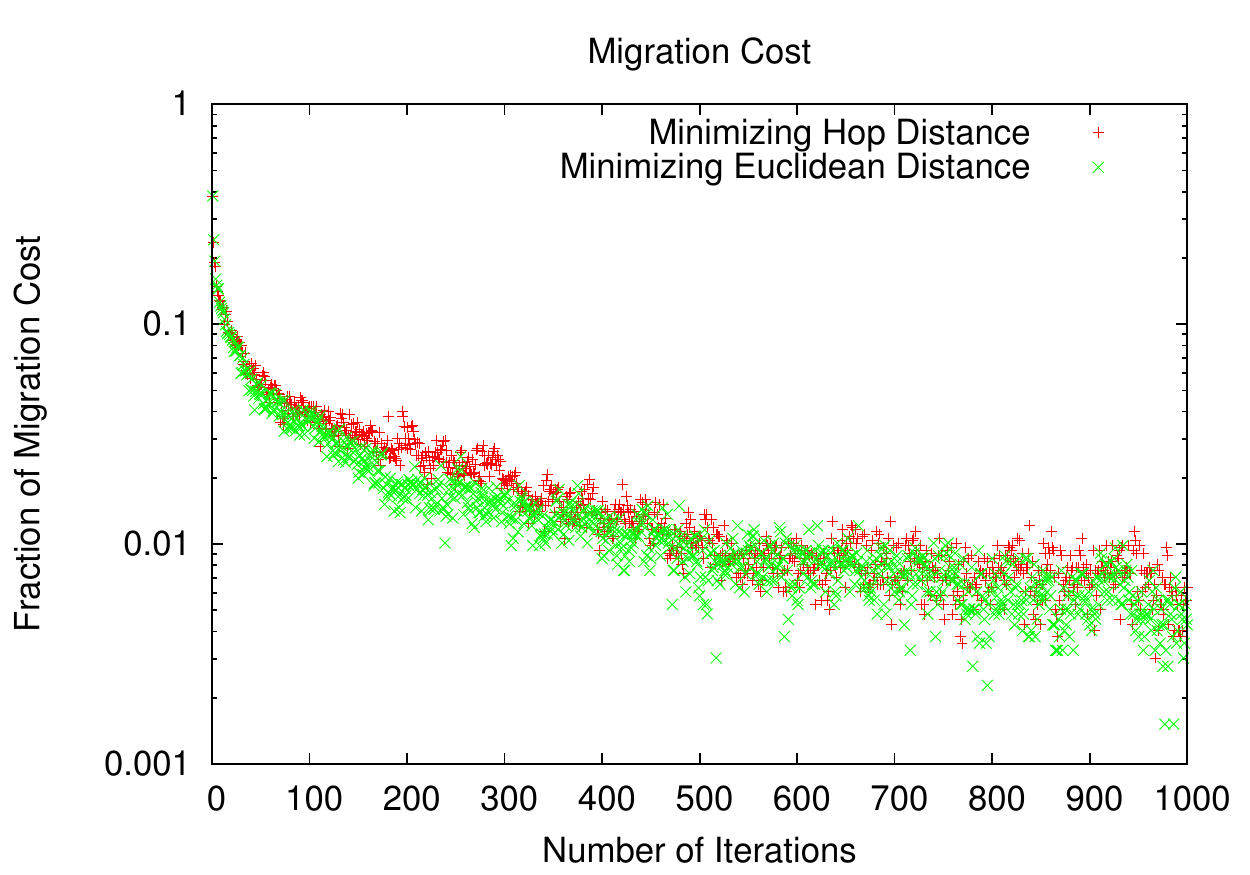}
\caption{Comparison of two different utility functions for the migration cost of the algorithm.
}
\label{fig:peer-sampling-swap}
\end{center}
\end{figure}

Figure \ref{fig:peer-sampling-swap} represents the migration cost of the algorithm, for two different cost functions, i.e., minimizing the number of hops and minimizing the euclidean distance.
Results show that both utility functions require similar migration cost.
Moreover, we observe low migration cost for the algorithm to achieve the improved performance observed in the previous result.
Based on the similarity in the cost and performance gain results, in the subsequent experiments we select the utility function of minimizing the euclidean distance due to it's simple decentralized nature.

\begin{figure}[t]
\begin{center}
\includegraphics[width=\columnwidth]{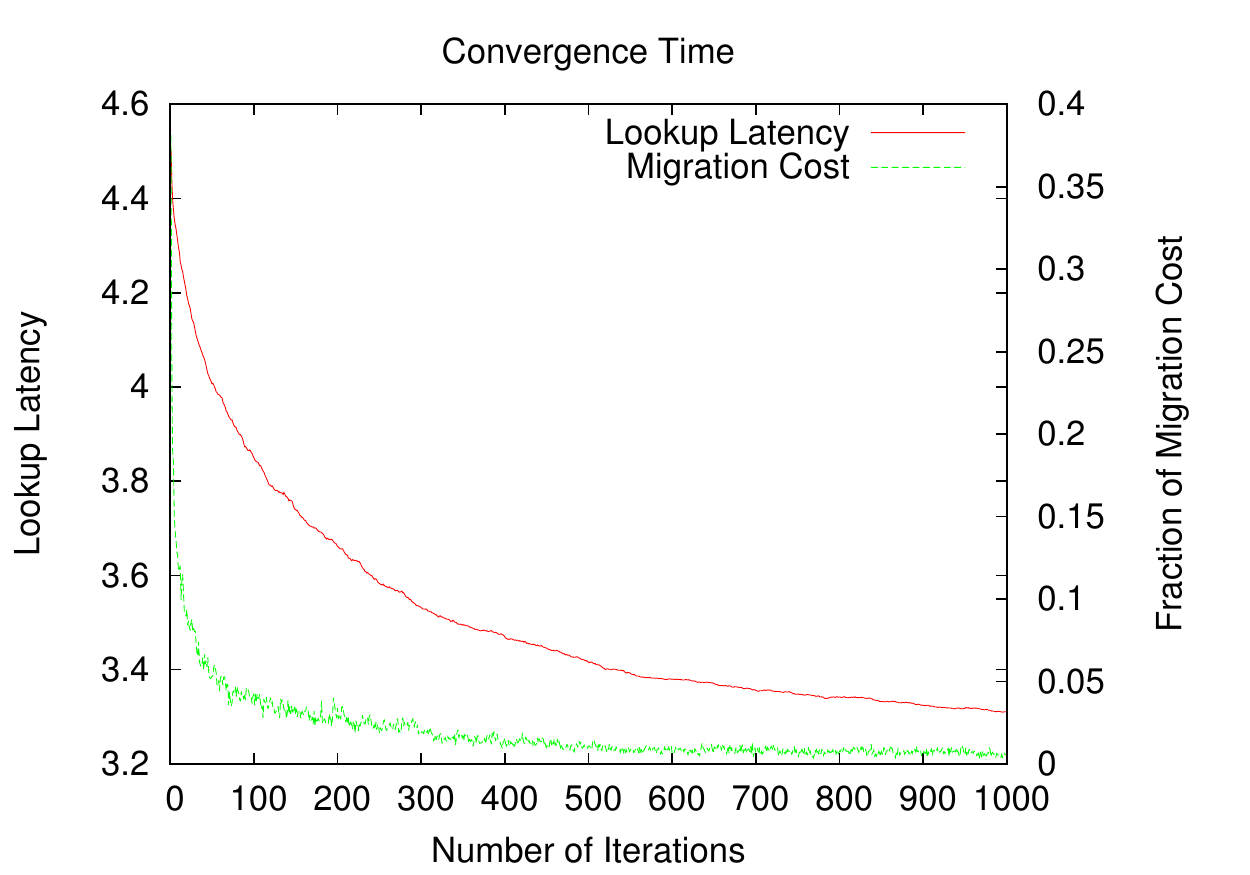}
\caption{Performance gain and migration cost of algorithm. Higher lookup gains come with higher migration cost.}
\label{fig:peer-sampling1}
\end{center}
\end{figure}

In the third experiment, we observe the convergence time of the algorithm, both in terms of migration cost and lookup latency.
Figure~\ref{fig:peer-sampling1} presents these results.
We observe that we achieve 30\% improvement in lookup latency by moving only 10\% of the nodes in the overlay.
Whereas, if we let the algorithm to run for a longer period, the average lookup latency improves to about 3.2 hops.
This gain is achieved by only moving few nodes in the overlay.
Another insight that we gain from this experiment is that very few nodes tend to move or exchange their identifiers: they quickly enter into a comfortable region in the overlay and stabilize their swaps.

\textbf{Q2:} As our algorithm is decentralized, each node in the overlay is capable of executing the algorithm on their own time.
An approach that mimics a realistic setting is to allow random nodes to perform the algorithm, i.e., initialization phase and refinement.

Other, more complicated techniques can first order nodes based on their degree centrality, which indicates importance in the network, and then use this order to select the next node to perform the algorithm.
On the one hand, top degree nodes are more important in the network and can affect many nodes in the overlay at once, thus, affecting its speed of convergence (could be faster, or may lead to oscillations).
On the other hand, bottom degree nodes are more periphery nodes and may allow slower, but steadier, convergence.

Such techniques are not easy to implement in a decentralized system: either a centralized service can provide the degree centrality ranking, or a decentralized service is needed to estimate global ranking.
However, such rankings are not easy to acquire in a decentralized setting; partial rankings can also be computed to elect nodes to start the algorithm.

\begin{figure}[t]
\begin{center}
\includegraphics[width=\columnwidth]{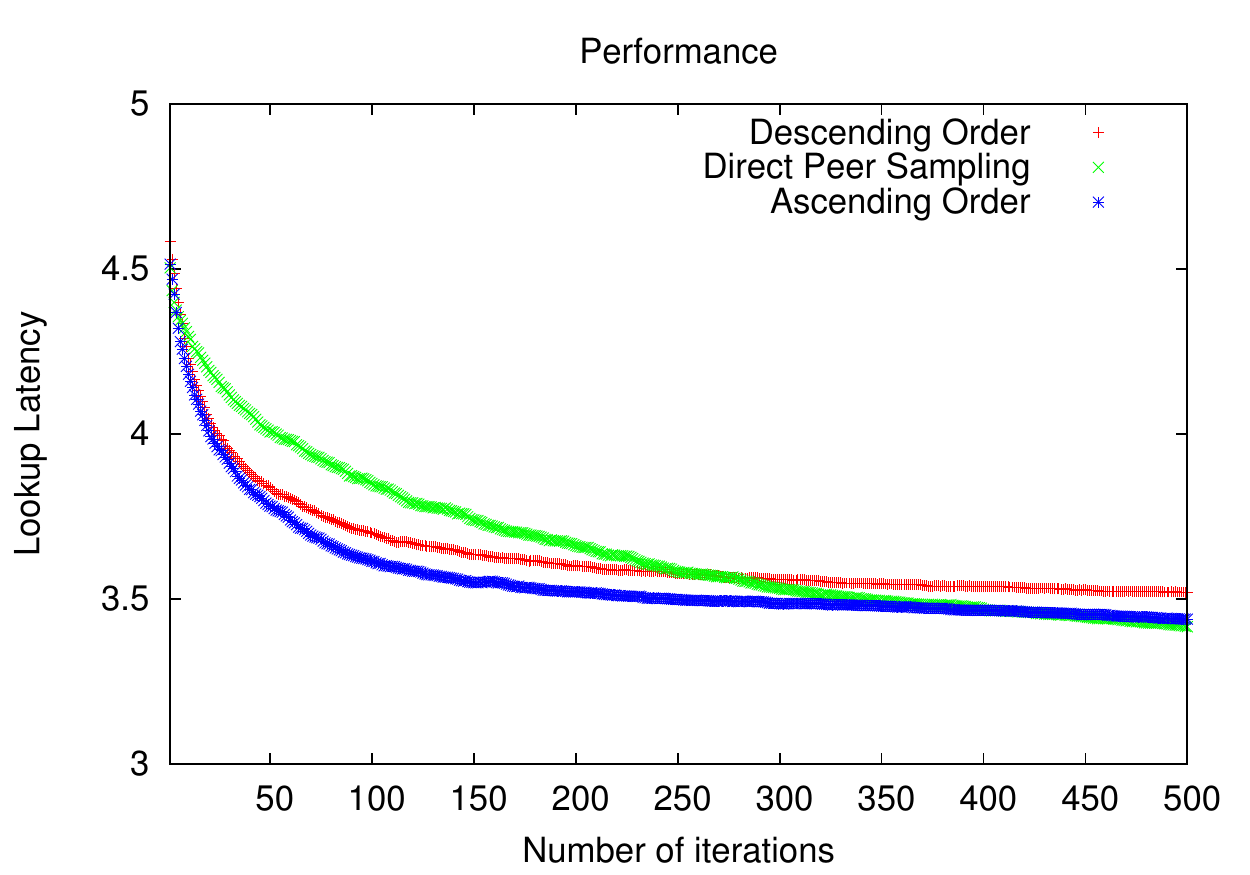}
\caption{Performance gains of three execution node orderings: a) descending order (execution starts from high degree nodes), b) direct peer selection, and c) ascending order (execution starts from low degree nodes).}
\label{fig:peer-performance-execution}
\end{center}

\end{figure}

\begin{figure}[t]
\begin{center}
\includegraphics[width=\columnwidth]{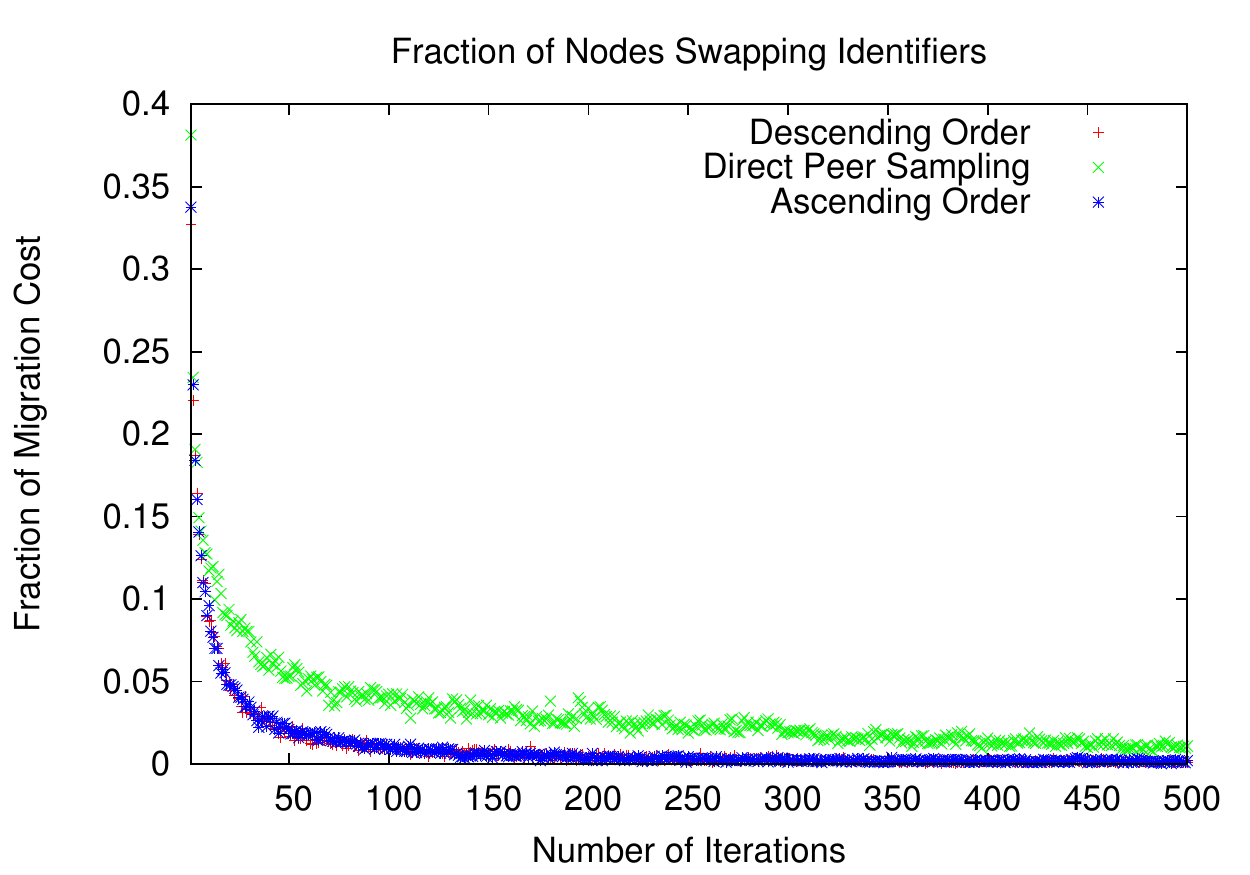}
\caption{Migration cost of three execution node orderings: a) descending order (execution starts from high degree nodes), b) direct peer selection, and c) ascending order (execution starts from low degree nodes).}
\label{fig:peer-swap-execution}
\end{center}
\end{figure}

We experiment with three different techniques to investigate if the additional complexity required by such techniques is granted with any additional performance gains.
In particular, we compare executions in a) descending order (starting from highest degree node), b) random order (through direct peer selection), and c) ascending order (starting from lowest degree node).
Figure \ref{fig:peer-performance-execution} and Figure \ref{fig:peer-swap-execution} show the performance gains and migration cost of the three approaches.
Compared to the direct peer selection scheme, ordering of nodes leads to faster convergence with lower migration cost, pointing to an interesting future direction of research.


\begin{figure}[ht]
\begin{center}
\includegraphics[width=\columnwidth]{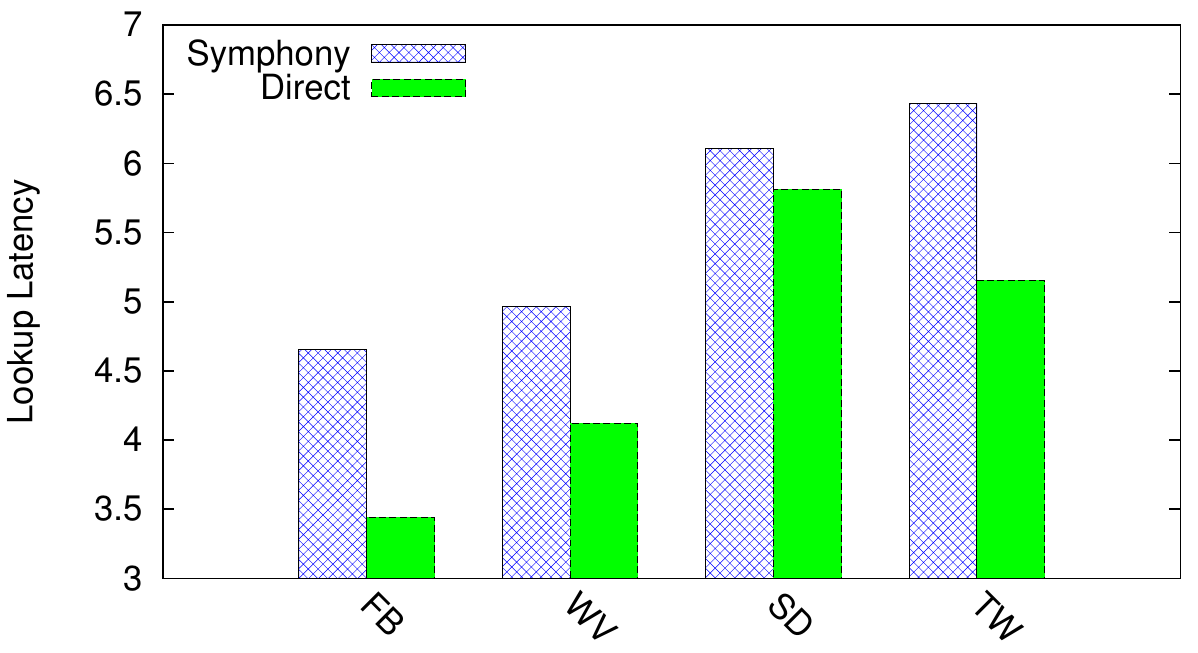}
\caption{Comparison of lookup latency using random and the gossip based overlay for different datasets. }
\label{fig:lookup-many}
\end{center}
\end{figure}

\textbf{Q3:} In this experiment, we evaluate the performance gain, i.e., lookup latency and reliability.
We run experiments for 500 iterations using different datasets (FB, WV, SD, TW).
Based on previous results, we use the direct peer selection and the utility function of minimizing the euclidean distance.
In the first experiment, we report the lookup latency for the symphony overlay and an overlay that is generated after running our algorithm.
The lookup latency represents the average communication hops that a node requires to lookup for their friends.
Figure \ref{fig:lookup-many} shows the different performance gains in latency for different datasets.
FB, WV, and TW datasets have higher gain in lookup latency compared to SD.
This behavior can be attributed to the low clustering in the SD dataset, which results in ambiguous segregation between social ties.
Moreover, the average degree in SD dataset is lower compared to other datasets, which gives less freedom of choice using the direct peer selection scheme, resulting in lower gains in performance. 

In the next experiment, we measure the reliability in the form of average gain.
We define two different metrics for average gain and perform experiments to measure both metrics using different datasets.
First metric calculates the reliability in terms of the finger table, which is given by the percentage of 1-hop friends versus the total number of fingers per node in the overlay:
\begin{center}
$reliability_1 \% = \dfrac{\text{\# of friends in the finger table}}{\text{\# of fingers}}$
\end{center}

\begin{figure}[ht]
\begin{center}
\includegraphics[width=\columnwidth]{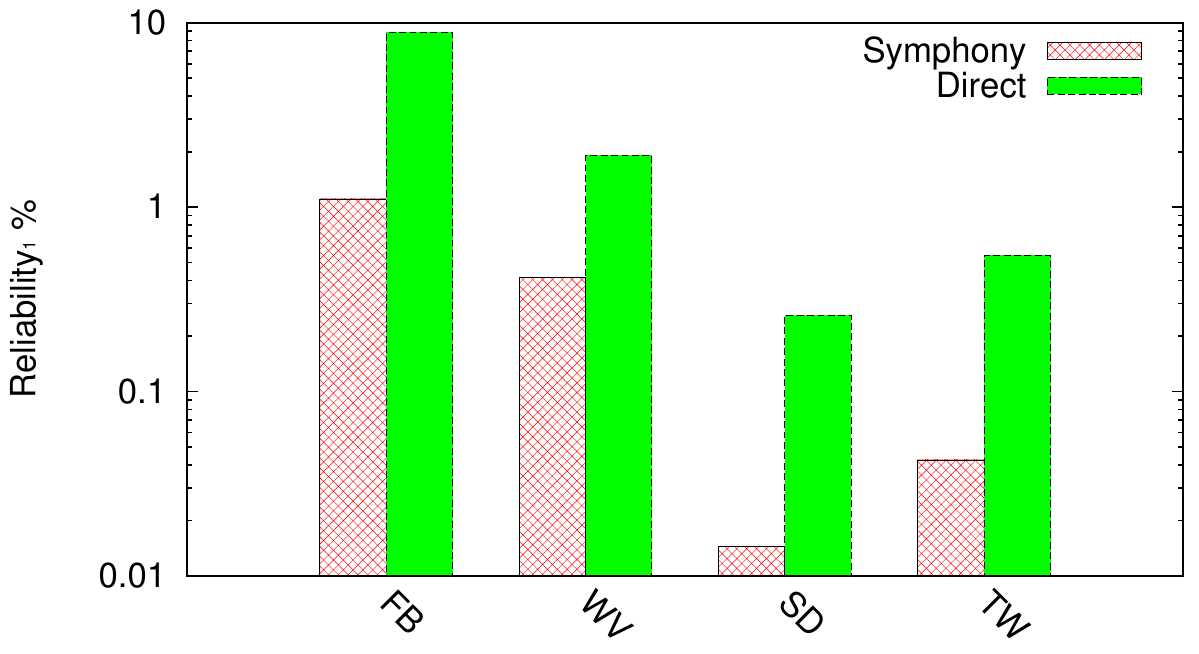}
\caption{Comparison of gain in reliability between symphony overlay and our approach using multiple datasets.}
\label{fig:reliability1}
\end{center}
\end{figure}

Figure \ref{fig:reliability1} compares the $reliability_1$ metric for a random symphony overlay and our approach.
In this experiment, we run our algorithm for 500 iterations using multiple datasets.
Results show high reliability gain in terms of the finger tables.
For instance, for FB, using random symphony overlay about 1\% of friends on average are in the finger table, whereas after running our algorithm, we have nearly 10\% of the fingers pointing to direct friends. 

Second metric evaluates the reliability in terms of social friends.
This is given by the percentage of 1-hop social friends positioned in the i-hop distance in the overlay, with $i \in \{1,2,3\}$.

\begin{center}
$reliability_2^i \% = \dfrac{\text{\# of friends in \textit{i}-hop in overlay}}{\text{\# of friends}}$
\end{center}

\begin{figure}[ht]
\begin{center}
\includegraphics[width=\columnwidth]{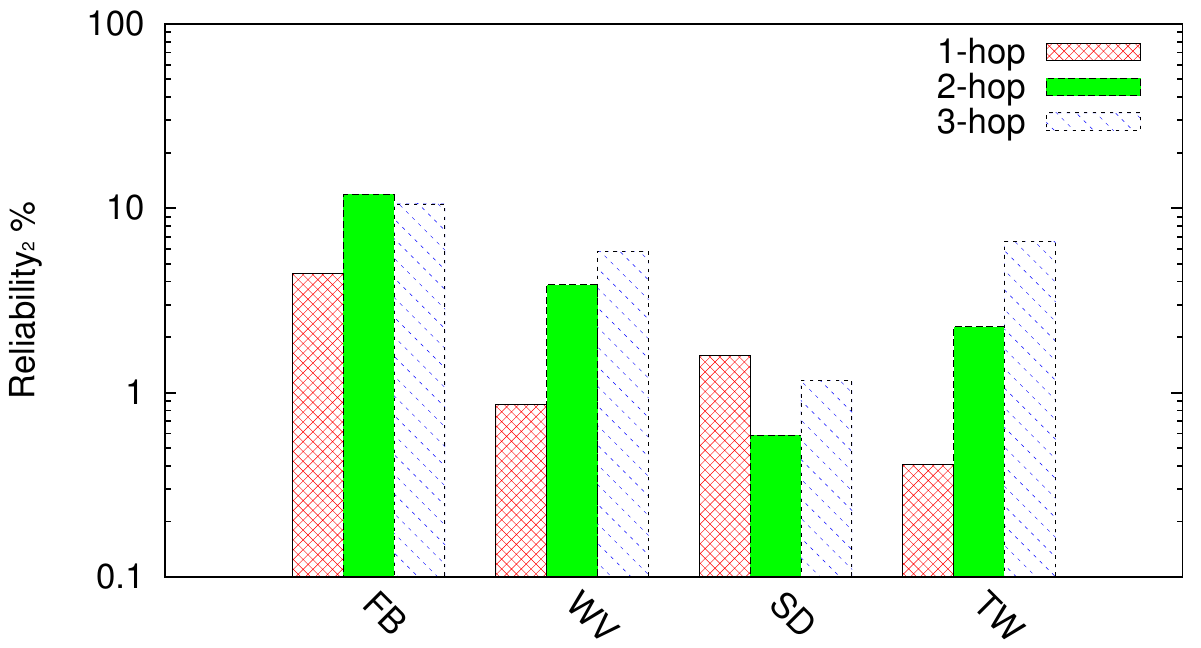}
\caption{Comparison of gain in reliability in terms of increase in 1-hop, 2-hop and 3-hop neighbors for multiple datasets. }
\label{fig:reliability2}
\end{center}
\end{figure}

Figure \ref{fig:reliability2} compares the $reliability_2$ metric for a random symphony overlay and our algorithm.
In this experiment, we run our algorithm for 500 iterations using multiple datasets.
We observe a significant increase in the amount of social friends in 1-hop, 2-hop and 3-hop, from the symphony overlay to an overlay that is generated after running our algorithm.
Therefore, we conclude that our algorithm is capable of placing users and their friends closer in the overlay with better lookup guarantees.
The gain in reliability enables using social friends in a more trustworthy lookup process. 

\textbf{Q4:} In this experiment, we study the robustness of the algorithm with respect to the clustering of the social graph.
In order to study this aspect of the algorithm, we begin by creating a symphony overlay with 10000 nodes \cite{manku2003symphony}.
We extract the fingers per node and construct a graph from these edges (step 1).
Then, we run our algorithm (step 2), that first initializes the new overlay by randomly assigning nodes from step 1 in the new DHT, and then refining their position in the new overlay by moving them closer to their neighborhoods from step 1.
The goal of this experiment is to measure how close we can move the nodes in the new overlay, to recreate the graph constructed from step 1 that comes with bounded latency of 1-hop.

In general, a symphony overlay from step 1 is expected to have lower number of triangles.
Therefore, triangle count might not be a good parameter to estimate the strength between nodes.
For such datasets with lower number of triangles, we need to rely on other parameters for social strength, like round trip time between nodes.
For this reason, we mimic the strength of ties using euclidean distance between node ids.

\begin{figure}[]
\begin{center}
\includegraphics[width=\columnwidth]{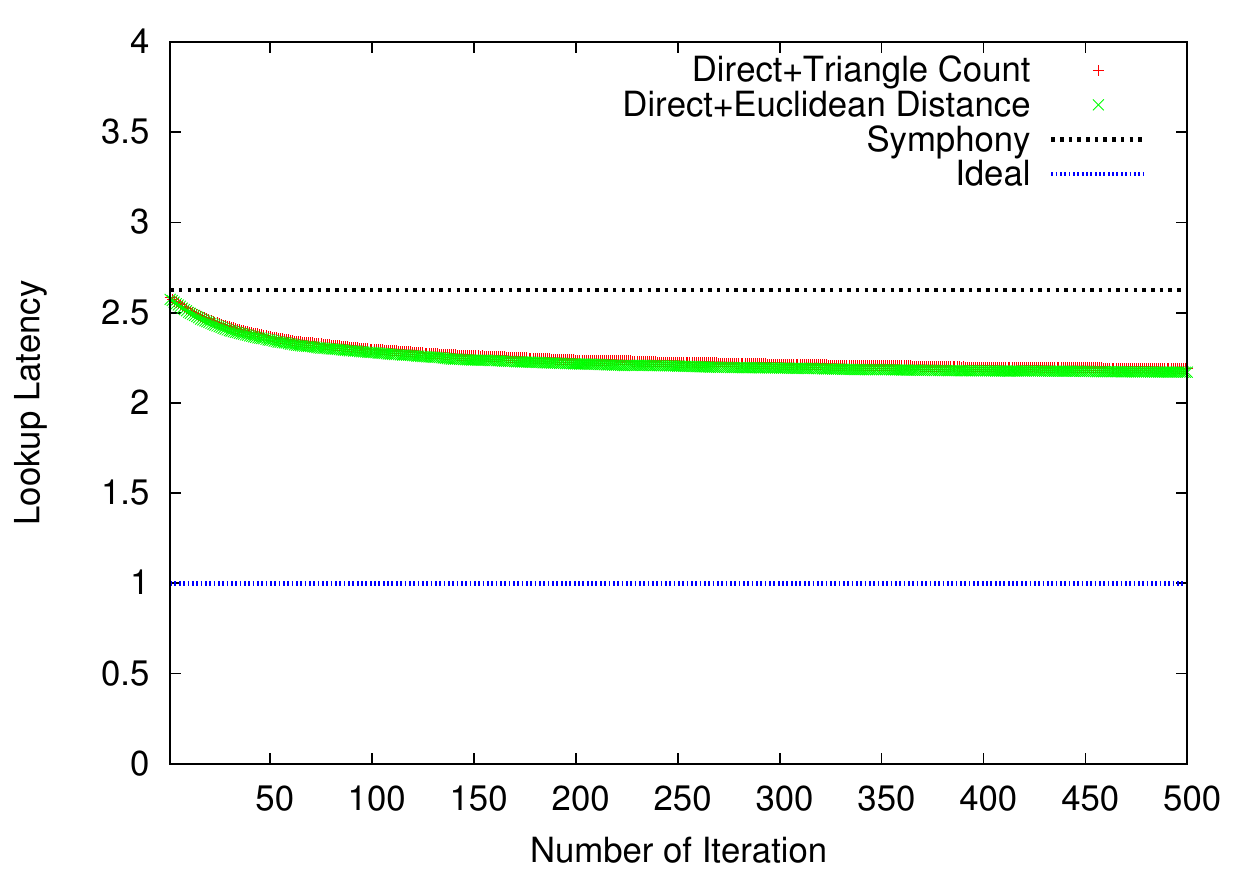}
\caption{Using the algorithm to relabel a symphony overlay. Comparison between different algorithm related parameters.
}
\label{fig:clustering}
\end{center}
\end{figure}

Figure \ref{fig:clustering} shows the comparison of these two strategies (direct+triangle count and direct+euclidean distance).
For reference, the plot also has a line for the ideal latency of 1-hop from step 1, and a line for the random overlay at the beginning of step 2. 
We achieve 22\% gain in latency compared to the Symphony lookup latency at the beginning of step 2.
Moreover, we observe a minor improvement using the euclidean distance as definition for strength between nodes.

%% file: discussion.tex
\section{Conclusion}\label{sec:discussion}

In our work, we proposed a socially-aware distributed hash table for decentralized online social networks.
In particular, we presented a gossip-based algorithm to place social users in a DHT, while maximizing the social-awareness among social users.
Furthermore, we perform several experiments with real graphs to evaluate the improvements of our proposal with respect to host access latency reduction and reliability improvements.
We believe that this approach will enable efficient and scalable implementation of various DOSN services.

There are various decentralized online social network services that benefit from the socially-aware distributed hash tables.
Such overlays ensure that users in social network reside ``close'' to their friends. 
The applications that take immediate benefits from these improvements include information dissemination, distributed storage, fault tolerance, publish/subscribe and others.

For example, information dissemination using socially-aware DHTs requires fewer hops for users to reach their friends in the overlay.
As shown in~\cite{marti2005dht}, friends of users are more likely to deliver or forward their messages compared to random nodes.
With our method, many of the friends can reside nearby in the overlay and thus, the overall communication cost can be decreased and the reliability of the system can be improved.
Furthermore, in a friend-to-friend storage system, the store and access latency of the data can be improved: fewer hops in the data transfers can mitigate connection failures and offer an overall faster service.

\section{Acknowledgment}
This work is supported by iSocial EU Marie Curie ITN project (FP7-PEOPLE-2012-ITN).
We would also like to thank all the anonymous reviewers for their constructive feedback.